\documentstyle[preprint,aps,epsfig]{revtex}
\begin{document}
\draft
\tighten
\preprint{FERMILAB-PUB-96/056-E}

\title{Properties of Jets in $Z$ Boson Events from 1.8 TeV
$\overline{p}p$ Collisions}

\author{
F.~Abe,$^{14}$ H.~Akimoto,$^{32}$
A.~Akopian,$^{27}$ M.G.~Albrow,$^7$ S.R.~Amendolia,$^{23}$ 
D.~Amidei,$^{17}$ J.~Antos,$^{29}$ C.~Anway-Wiese,$^4$ S.~Aota,$^{32}$
G.~Apollinari,$^{27}$
% ...
\\
(CDF Collaboration)
}
\address{
% ...
$^4$  {University of California at Los Angeles, Los 
Angeles, California  90024} \\  
% ...
$^7$  {Fermi National Accelerator Laboratory, Batavia, Illinois 
60510} \\
% ...
$^{14}$ {National Laboratory for High Energy Physics (KEK), Tsukuba, 
Ibaraki 305, Japan} \\
% ...
$^{17}$ {University of Michigan, Ann Arbor, Michigan 48109} \\
% ...
$^{23}$ {Istituto Nazionale di Fisica Nucleare, University and Scuola
               Normale Superiore of Pisa, I-56100 Pisa, Italy} \\
% ...
$^{27}$ {Rockefeller University, New York, New York 10021} \\
% ...
$^{29}$ {Academia Sinica, Taipei, Taiwan 11529, Republic of China} \\
% ...
$^{32}$ {University of Tsukuba, Tsukuba, Ibaraki 305, Japan} \\
% ...
}

\date{\today}
\maketitle
\begin{abstract}
We present a study of events with $Z$ bosons and hadronic jets 
produced in $\overline{p}p$ collisions at a center-of-mass 
energy of 1.8 TeV. The
data consist of 6708 $Z \rightarrow e^+e^-$ decays 
from 106 pb$^{-1}$ of integrated luminosity
collected using the CDF detector at the Tevatron Collider. The
$Z$ $+ \ge n$ jet cross sections and jet production properties 
have been measured for $n=1$ to 4.  The data compare well 
to predictions of leading order QCD matrix element calculations
with added gluon radiation and simulated parton fragmentation. 
\end{abstract}


    The $Z$ boson was discovered in $\overline{p}p$ collisions at CERN in 
1983\cite{Z_1,Z_2}. Since
then, the hadronic production properties of the $Z$ have been studied using
relatively small event samples, most recently from experiments at the Fermilab 
Tevatron Collider\cite{zrecent}. In this Letter we describe an
extension of these studies based on a much larger sample of 
6708 $Z \rightarrow e^+e^-$ decays
obtained from 106 pb$^{-1}$ of integrated $\overline{p}p$  
luminosity using the CDF detector. This event sample is 
large enough to study the production
properties of high energy hadronic jets associated with $Z$ boson production.
The $\overline{p}p \rightarrow Z$ + jet events provide a good 
test of Quantum Chromodynamics (QCD) calculations
since the event sample has very small background and
the presence of the $Z$ selects high $Q^{2}$ parton-level 
processes.  A determination of the reliability of QCD calculations for heavy
boson production is important for verifying the direct $W$ boson
backgrounds to top quark production\cite{topprd,topprl}.  
$Z$ boson events are free of top
quark contamination and provide a clean test of Standard Model heavy boson
production.  As part of this study we examine the $Z$ boson sample for 
evidence of excess $b$ quark decays that could indicate
new particle production.

    The elements of the CDF detector of primary importance to this
analysis are the central tracking chamber (CTC), the calorimeters, and the
silicon vertex detector (SVX).  The
CTC, which is immersed in a 1.4 T solenoidal magnetic field, measures the
momenta and trajectories of charged particles in the region
$|\eta| < 1.1$ (where $\eta = - \ln (\tan (\theta/2))$\cite{coord}).
The calorimeters are divided into electromagnetic and hadronic
components and cover the pseudorapidity range $|\eta| < 4.2$.  
The four-layer silicon strip detector (SVX)\cite{svx}, 
located just outside the beampipe,
provides precise
tracking in the plane transverse to the beam direction and 
is used to reconstruct secondary decay vertices from $B$ hadrons. 
The CDF detector is described in detail elsewhere\cite{topprd,CDF_Detectors}.

    Events with $Z$ bosons are identified by looking for the decay
$Z \rightarrow e^+e^-$.  
We demand that candidate events pass a high-$E_T$ electron trigger.
The event selection requires an isolated electron\cite{iso} 
that satisfies tight
selection cuts\cite{tight_cuts} and has a track in the central 
region ($|\eta|\le 1.1$). Events
containing a central electron with 
transverse energy $E_T \ge 20$ GeV are required
to have a second electron that satisfies looser selection cuts and has
opposite charge where the sign of the track curvature is well-measured.
The second 
electron is detected in either the central calorimeter 
($E_T \ge 20$ GeV and $|\eta|\le 1.1$), the plug calorimeter
($E_T \ge 15$ GeV and $1.1 \le |\eta|\le 2.4$), or the forward
calorimeter ($E_T \ge 10$ GeV and $2.4 \le |\eta|\le 3.7$). 
We remove electrons with photon conversion characteristics.
The separation in $\eta-\phi$ between the centroid of an electron 
calorimeter cluster and that of any jet in the event, measured in
$\Delta R = \sqrt{ \Delta\eta^2 + \Delta\phi^2}$, must exceed
 $\Delta R = 0.52$.

      An event sample of 6708 $Z \rightarrow e^+e^-$ 
decays is selected  by requiring the
electron pair masses to be within 15 GeV/c$^2$ of the nominal 
$Z$ boson mass of 91 GeV/c$^2$.
The background in these data and the $Z \rightarrow e^+e^-$ 
acceptance due to our 
selection cuts are measured 
as a function of associated jet multiplicity as described below.

    Hadronic jets produced in association with the $Z$ bosons are selected 
using a clustering algorithm\cite{jetclu} with a cone 
size of $R_{j}=0.4$.  The jet energies are corrected 
to account for variations in calorimeter response, estimates of 
fragmentation energy 
outside the jet cone, and underlying event energy within the jet cone. 
Initially, we consider all jets with corrected transverse energy $E_T \ge$ 12 
GeV.  When the separation between two jets is less than $\Delta R = 0.52$, 
they are combined vectorially into a
single jet.  The requirements $E_T \ge 15$ GeV and $|\eta|\le 2.4$ 
are applied to obtain the hadronic jets used for this analysis.
In our sample of 6708 $Z$ events, 1310 have $\geq$1 jet, 279 have 
$\geq$2 jets, 57 have $\geq$3 jets, and 11 have $\geq$4 jets.
We correct these jet multiplicities for two additional effects:  photons
counted as hadronic jets, and jets produced in extra $\overline{p}p$ 
interactions that occur in the same bunch crossing as the $Z$ event. A
$Z$ $+$ photon Monte Carlo calculation\cite{BaurZMC} with detector
simulation yields a correction varying from --2\% to --3\% for photons
as a function of jet multiplicity.  Using minimum bias events, we estimate
the number of jets from additional interactions that pass the selection cuts,
and obtain corrections varying from --3\% to --5\%.

    The backgrounds to the $Z$ boson are dominated by jets faking electrons,
but include some contributions from heavy quark, 
$W \rightarrow e\nu$ + jet, and $Z \rightarrow \tau^+\tau^-$ decays.
To measure the background we employ a data sample in which all $Z$ boson
selection cuts have been applied, except the mass window and
electron isolation cuts.  By selecting events from this sample in
which neither electron candidate is isolated, we obtain a set of events
that is almost entirely background with no measurable
contribution from $Z \rightarrow e^+e^-$.  The mass distribution for these
background events is independent of electron candidate isolation.  This
allows us to estimate the number of background events with
isolated electrons that lie within the mass window.  
This method yields background estimates in
the $Z$ boson event sample that are small even at high jet multiplicities. 
The $1 \sigma$
upper limits for the backgrounds are 1.1\%, 2.3\%, 3.0\%, and 4.0\% for the 
$n \geq$ 0, 1, 2, and 3 jet events, respectively. 


    The acceptance for the electrons from a $Z$ boson decay to pass all
selection cuts has been measured as a function of the number of 
associated hadronic jets. Losses due to the electron $E_T$ cuts and 
detector geometric acceptance requirements have been studied 
using an inclusive $Z$ boson production model
and a leading-order $Z$ + jets QCD calculation \cite{VECBOS}.
The loss of electrons due to overlap with hadronic jets is
determined directly from the data by taking $Z \rightarrow e^+e^-$ 
events and using a Monte
Carlo program to re-decay the $Z$ bosons.  
The efficiency for decay electrons to be separated from jets
and to be isolated sufficiently from other energy in the calorimeters 
is determined as a function of jet multiplicity. The systematic 
uncertainties in the overlap efficiencies are determined
by varying the polarization of the $Z$ and
the separation between the electrons and any jets in the event.
For electrons passing the $E_T$, geometric acceptance, and electron-jet
separation cuts, a final acceptance correction is made for
electron identification cuts and the efficiency of the online
trigger, which selects a high-$E_T$ central electron. 
The total $Z \rightarrow e^+e^-$ 
detection efficiencies vary from $(37.3\pm0.7)\%$ 
for $Z$ bosons without
jets to $(31.5\pm1.4)\%$ for $Z$ bosons with $\ge4$ jets.

    The systematic uncertainties on the number of jets due to the jet selection
cuts are determined by varying the jet energy scale by $\pm$5\%, the $|\eta|$
cut by $\pm$0.2, the underlying event correction by $\pm$50\%, the photon
removal correction by $\pm$15\%, and the probability of jets from
additional $\overline{p}p$ interactions by $\pm 100$\%. The total jet-counting
uncertainties are found to range between 11\% for the $\geq 1$ jet to 23\% for
the $\geq 4$ jet sample.  These errors dominate the uncertainties in the cross
section measurements, except for $Z$ $+ \geq 4$ jet events where the
statistical error is larger.
    
    We measure the cross section for $Z$ production as a function of jet
multiplicity using the ratio of the number of detected
$Z$ events with $\geq n$ jets to the total number of detected $Z$ events.  
The cross sections are then determined from this ratio using 
the inclusive $Z$ boson cross section of $231\pm 12$ pb \cite{xsec}
and the ratios of detection efficiencies.  This method takes advantage of the
cancellation of some systematic uncertainties in the ratios, and gives the most
accurate relative $Z$ $+ \geq n$ jet cross sections.  The measured cross
sections for $Z$ $+ \ge n$ ($n=1$ to 4) jet events are given in 
Table~\ref{tab:results}, as are the measured ratios of
$\sigma(Z + \ge n$ jets$)/\sigma(Z + \ge n-1$ jets), which are constant at
$\sim$0.2 for $n=1$ to 4.

     The measured cross sections and the production characteristics of
the hadronic jets can be compared to QCD calculations. The leading-order matrix
element calculations for $Z + n$ parton events 
(for $n = 1$,~2,~3) are obtained with the
VECBOS\cite{VECBOS} Monte Carlo program\cite{parton_range}. We use both
MRSA and CTEQ3M parton distribution functions, the two-loop $\alpha_{s}$
evolution, and factorization and renormalization scales varying from  
$Q^{2} = M_{Z}^{2} + p_{T_{Z}}^{2}$ of the $Z$ boson to
$Q^2 = \langle p_T \rangle^{2}$, where $\langle p_T \rangle$ is the average
$p_T$ of the generated partons.
The QCD predictions are indistinguishable for the two parton
distribution functions within the statistical uncertainties of our 
calculation.

    The QCD-predicted $Z$ + jet events are obtained from the parton-level  events
by including gluon radiation and hadronic fragmentation using the
HERWIG\cite{herwig,herprt} shower simulation algorithm\cite{ISradiation}. This
procedure represents a partial higher-order correction to tree-level
diagrams. The $Z$ boson 
events with hadron showers are then introduced into a full CDF
detector simulation,  and the resulting jets are identified and selected  in a
manner identical to the data. This procedure allows us to make direct 
comparisons between the QCD predictions and 
data.

\begin{figure}[htb]
\begin{center}
\mbox{\epsfig{file=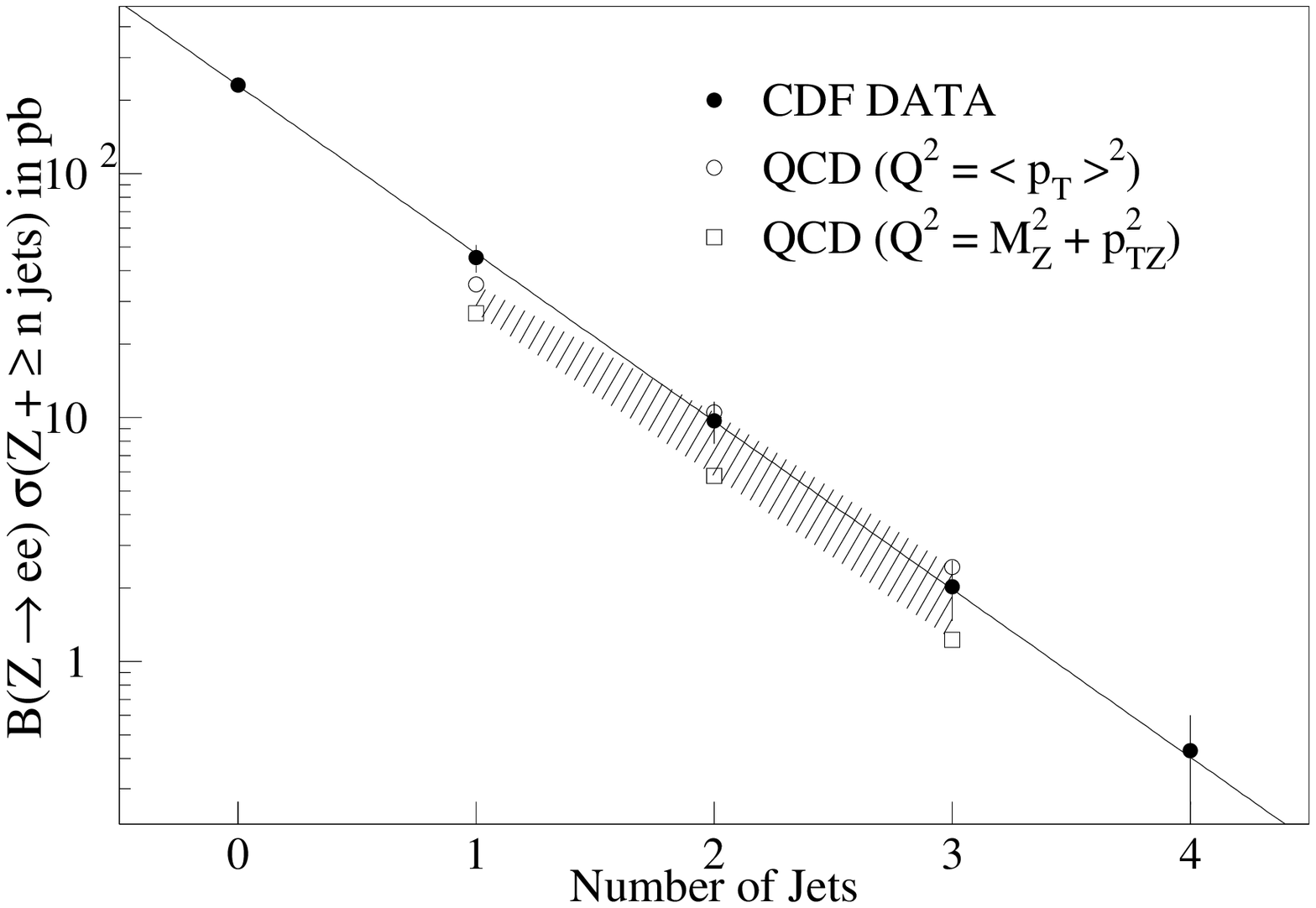,width=6.4in}}
\nobreak
{\small \caption{
Cross section for $Z$ $+ \geq n$ jets versus $n$, where $Z \rightarrow e^+e^-$.
The QCD prediction band is determined by using two different renormalization
scales.  The solid line is an exponential fit to the measured cross sections.
The error bars include statistical and systematic uncertainties.
\label{fig:xsec}}}
\end{center}
\end{figure}

The Monte Carlo events passing our jet selection cuts are used to determine
QCD-predicted cross sections that are compared to the experimental
measurements corrected for $Z \rightarrow e^+e^-$ decay losses and 
backgrounds.  These calculated QCD cross sections are given in 
Table~\ref{tab:results}, and compared to the data via ratios of the
measured to QCD-predicted cross sections.  Figure~\ref{fig:xsec} shows a
plot of the measured cross sections for  $Z$ $+ \ge n$ jets as a function
of $n$, with the QCD predictions resulting from variations in $Q^2$
indicated by a superimposed band.  For  $Q^2 = \langle p_T \rangle^{2}$,
the measured $Z$ $+ \ge n$ jet cross sections range from 0.83 to 1.29 times the
leading-order QCD predictions. For $Q^{2} =
M_{Z}^{2} + p_{T_{Z}}^{2}$, the ratios of cross sections to leading-order QCD
predictions for each $n$ are larger, but nearly constant at 1.7.

\begin{figure}[htb]
\begin{center}
\mbox{\epsfig{file=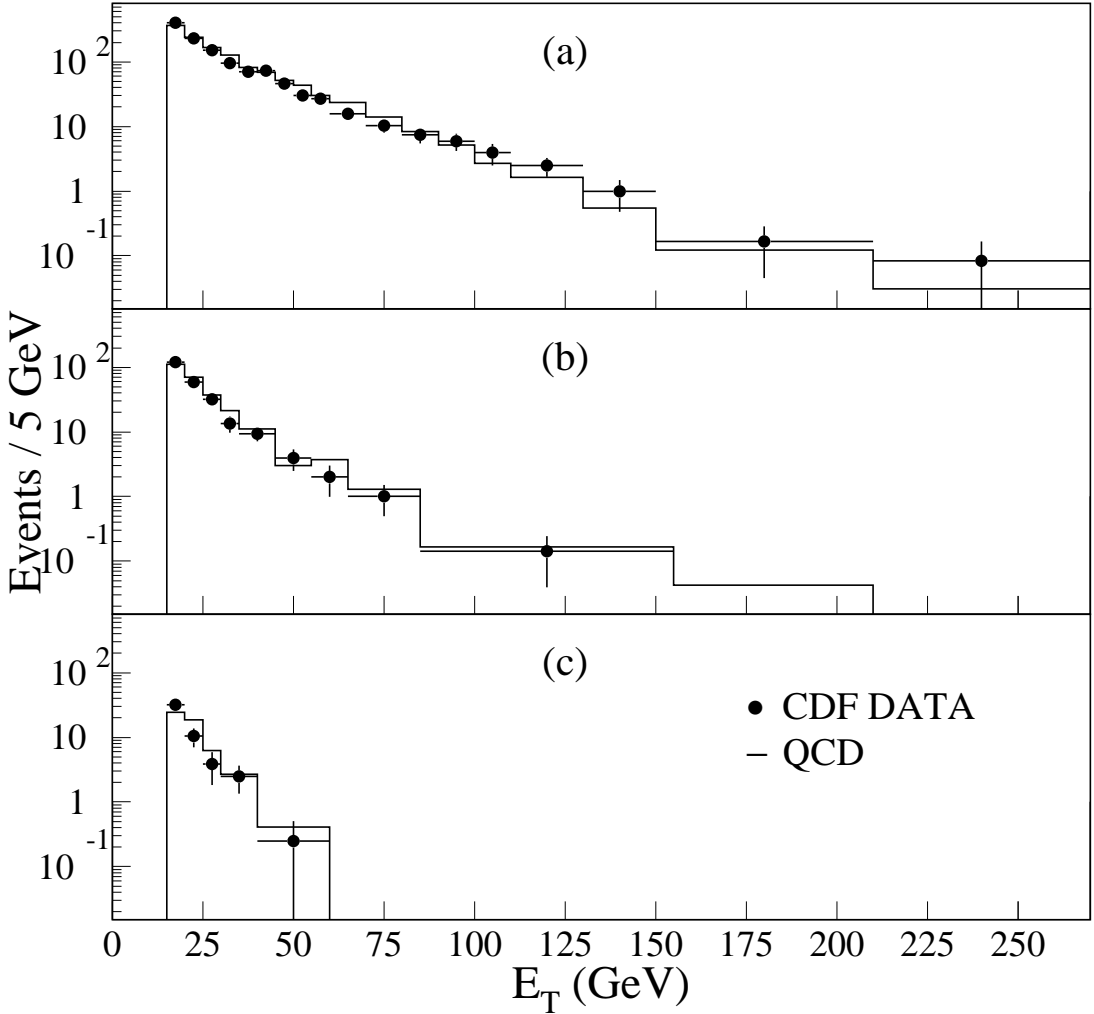,width=6.4in}}
\nobreak
{\small \caption{
Transverse energy of the (a) first, (b) second, and (c) third highest $E_T$ 
jets in $\geq$1, $\geq$2, and $\geq$3 jet events, respectively.  The points are
the data with statistical errors only, and the solid histograms are the
QCD predictions (normalized to the data) described in the text.
\label{fig:Et}}}
\end{center}
\end{figure}

\begin{figure}[htb]
\begin{center}
\mbox{\epsfig{file=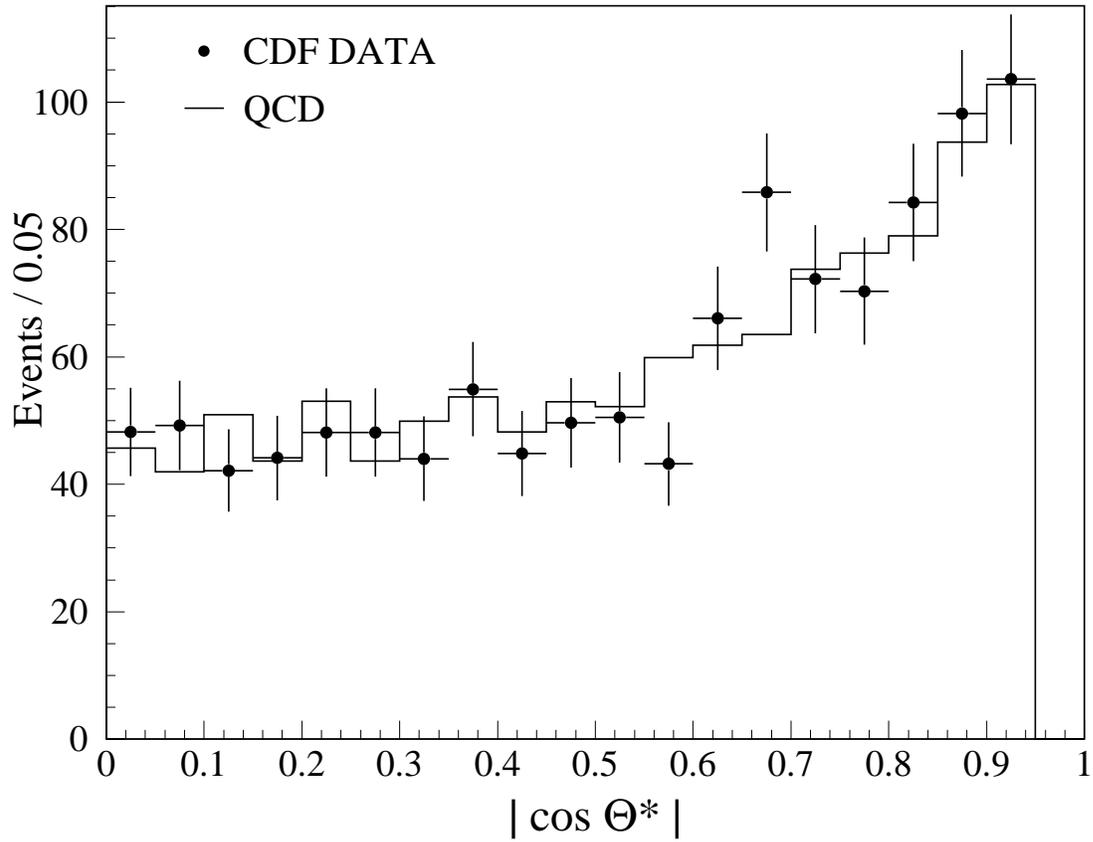,width=6.4in}}
\nobreak
{\small \caption{
The $|\cos \Theta^{*}|$ distribution of the $Z$ in $\ge$1 jet events.  
The angle $\Theta^{*}$ is the angle 
of the $Z$ measured with respect to the average beam direction 
in the $Z$ + leading jet center-of-mass frame. 
The Monte Carlo predictions are normalized to the data, and the
data errors are statistical only. 
\label{fig:cthstar}}}
\end{center}
\end{figure}

\begin{figure}[htb]
\begin{center}
\mbox{\epsfig{file=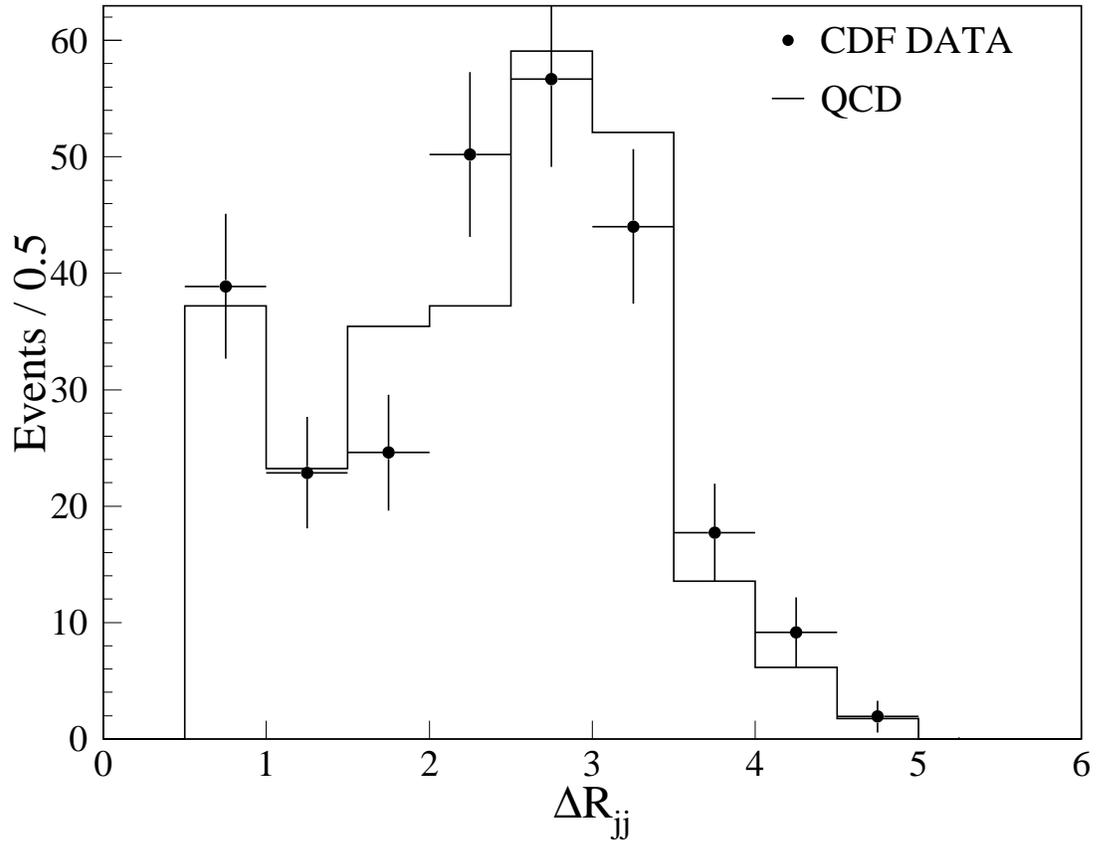,width=6.4in}}
\nobreak
{\small \caption{
Separation in $\eta-\phi$ space between the two leading jets
in $Z$ $+ \geq 2$ jet events.  The Monte Carlo predictions are
normalized to the data.  The data errors are statistical only.
\label{fig:drjj}}}
\end{center}
\end{figure}

      In comparing the jet production properties of $Z$ + jets events,  we use
the same  Monte Carlo event generation described above but apply both 
$Z \rightarrow e^+e^-$ and jet selection cuts. 
For this study the QCD-predicted event distributions with $Q^{2}$ =
$\langle p_T \rangle^{2}$ are normalized to the number of  events in the data
samples.  Corrections are made for photons counted as
hadronic jets and for jets from extra  $\overline{p}p$ 
interactions in the same crossing as the $Z$ event, but no corrections are made 
for  the very small $Z$ boson backgrounds.  Figure~\ref{fig:Et} shows the
$E_T$ spectra  for the first, second, and third jets (ordered by decreasing
$E_T$)  in the $\ge 1$, $\ge 2$, and $\ge 3$ jet $Z$ events, respectively.
In the $Z$ $+ \geq 1$ jet event sample we measure the angle 
$\Theta^{*}$  between the $Z$ boson and
the average beam direction in the $Z$ + leading jet center-of-mass frame. 
The distribution of the quantity $|\cos \Theta^{*}|$, for the range
$|\cos \Theta^{*}| < 0.95$, is shown in
Figure~\ref{fig:cthstar}.  In events with two or more jets, we measure the
separation in $\eta-\phi$ space  between the two leading jets.  The  
resulting $\Delta R_{jj}$ distribution is given in Figure~\ref{fig:drjj}.  
This distribution is not corrected for the small number of photons counted as 
hadronic jets.

      The measured jet production properties agree well with the QCD
predictions. The $\chi^2$ per degree of freedom is $16.1/18$ for the 
$\cos \Theta^{*}$
distribution shown in Figure~\ref{fig:cthstar} and $7.8/7$ for the $\Delta R_{jj}$
distribution shown in Figure~\ref{fig:drjj}. The QCD-predicted $\Delta R_{jj}$
distribution is sensitive to the manner in which gluon radiation is added by
the HERWIG algorithm to the VECBOS matrix element, and limiting this radiation
results in poorer agreement with the data. The $E_T$ distributions shown in
Figure~\ref{fig:Et} are also in reasonable agreement with the QCD predictions.
The values of the $\chi^2$ per degree
of freedom are $19.8/12$, $3.7/4$, and $6.2/2$ for the jet $E_T$ spectra shown
in Figures~\ref{fig:Et}a, \ref{fig:Et}b, and \ref{fig:Et}c, respectively. 

    The $Z$ + jets data sample has been 
examined for secondary vertices that are
characteristic of $b$ quark decays.  The secondary vertices are
measured using the SVX detector and are reconstructed using
the same secondary vertex algorithm developed by CDF and 
used in the top quark search\cite{topprd,topprl}.
Of the 1665 jets in the sample, 442 are candidates for secondary vertex
reconstruction, which means that they have at least two tracks in the SVX,
an uncorrected jet $E_T > 15$ GeV, and  $|\eta| < 2$.  A control sample of
inclusive jet events is used to determine the number of secondary
vertices expected from jets
in QCD events with no heavy boson.  In all, six secondary vertex
candidates are found in the data
sample where 6.3 $\pm$ 1.0 are expected, 
so there is no excess of $b$ quark decays over
the expected number in this $Z$ + jets dataset. 

In summary, this Letter contains an analysis of jet  production properties
associated with $Z \rightarrow e^+e^-$ events  
selected from 106 pb$^{-1}$ of  $\overline{p}p$ collisions
at a center-of-mass energy of 1.8~TeV.  Comparisons are made between the data
and leading order parton matrix elements with HERWIG-simulated 
parton shower and fragmentation.  The ratios of the measured
to QCD-predicted cross sections are found to vary from $1.29 \pm 0.17$ to $1.69
\pm 0.22$ for $\overline{p}p \rightarrow Z + \ge 1$ jet when the $Q^2$ scale is
varied from $\langle p_T \rangle^{2}$ to  $M_Z^{2} + p_{T_Z}^2$.  The
QCD-predicted jet production properties are in generally good agreement with
the measured distributions. The incidence of $b$ quark decays in the jets
associated with $Z$ bosons is consistent with that observed in similar jets from
events with no heavy boson. 

     We thank the Fermilab staff and the technical staffs of the participating
institutions for their vital contributions. We also thank Walter Giele and the
theory group at the University of Durham for many useful discussions.   This
work was supported by the U.S. Department of Energy and National Science
Foundation; the Italian Istituto Nazionale di Fisica Nucleare; the Ministry of
Education, Science and Culture of Japan; the Natural Sciences and Engineering
Research Council of Canada; the National Science Council of the Republic of
China; and the A. P. Sloan Foundation.

\widetext
\begin{table}[htb]
\begin{center}
\begin{tabular}{|c|c|c|c||c|c|c|}
 $n$ &
 $BR \times \sigma_{\rm Data}$ &
 \multicolumn{2}{c||}{$Q^2 = \langle p_{T} \rangle^2$} &
 \multicolumn{2}{c|}{$Q^2 = M_{Z}^2 + p_{T_{Z}}^2$} &
 $\sigma(n)/\sigma(n-1)$ \\
\cline{3-6}                   
 Jets &
 (pb) &
 $BR \cdot \sigma_{\rm QCD}$(pb) &
 $\sigma_{\rm Data}/\sigma_{\rm QCD}$ &
 $BR \cdot \sigma_{\rm QCD}$(pb) &
 $\sigma_{\rm Data}/\sigma_{\rm QCD}$ &
 (Data) \\
\hline
\hline

$\geq$ 0  &   $231  \pm 6 \pm 11$  
& ---               &  ---
& 223  $\pm$ 10   &   1.04 $\pm$ 0.07  
& ---               \\
\hline
\hline

$\geq$ 1  &   45.2  $\pm$ 1.2  $\pm$ 5.7  
&  35.16 $\pm$ 0.54  &  1.29 $\pm$ 0.17
&  26.82 $\pm$ 0.40  &  1.69 $\pm$ 0.22 
&  0.196 $\pm$ 0.007 \\
\hline

$\geq$ 2  &   9.7  $\pm$ 0.6  $\pm$ 1.8  
&  10.53 $\pm$ 0.38  &  0.92 $\pm$ 0.19
&   5.77 $\pm$ 0.16  &  1.68 $\pm$ 0.34
&  0.215 $\pm$ 0.014 \\
\hline

$\geq$ 3  &   2.03  $\pm$ 0.28  $\pm$ 0.49  
&  2.44 $\pm$ 0.17  &  0.83 $\pm$ 0.24
&  1.23 $\pm$ 0.08  &  1.66 $\pm$ 0.47
&  0.210 $\pm$ 0.027 \\
\hline

$\geq$ 4  &   0.43  $\pm$ 0.13  $\pm$ 0.11  
&    & 
&    &  
&  0.211 $\pm$ 0.059 \\
\end{tabular}
\smallskip
\caption{
$Z$ $+ \geq n$ jet Cross Sections.  The first error on the data cross
sections is the statistical error;
the second includes the systematic error on the $Z$ acceptance and the
luminosity error added in quadrature.  For $n=1$ to 4, the second error also
includes the jet-counting uncertainty, as described in the text. 
The leading order (LO) Monte Carlo cross sections are
generated with VECBOS for $Q^2$ scales of $\langle p_{T} \rangle^2$
and $M_{Z}^2 + p_{T_{Z}}^2$.  
The $\geq 0$ jet calculation is next-to-next-to-leading order (NNLO) with 
$Q^2=M_Z^2$.  The error on the NNLO QCD cross section includes systematic
and statistical uncertainties; for the LO cross sections the errors are
statistical.
\label{tab:results}}
\end{center}
\end{table}


\begin{thebibliography}{10}
\bibitem{coord}
We use a polar coordinate system in which $z$ is along the proton direction,
$\phi$ is the azimuthal angle, and $\theta$ is the polar angle.
Pseudorapidity ($\eta$)
is calculated using $z$~$=$~0 at the center of the detector for fiducial
cuts and 
at the interaction point for event variables (i.e.\ $\Delta R$ cones).
\bibitem{iso}
An isolated electron is defined to be one for which the transverse calorimeter
energy in the cone of radius 0.4 in $\eta-\phi$ space around the electron
cluster is less than 10\% of the electron transverse energy.
\bibitem{parton_range} We generate $Z$ + parton events with parton 
$p_T > 8$ GeV/c, $|\eta|< 3.5$, and $\Delta$R $> 0.4$ between partons.
\bibitem{ISradiation} 
The HERWIG initial state radiation
contributes jets to the $Z$ boson event.  For the purpose of the calculation
presented here, a kinematic limit of $\sqrt{M_{Z}^{2} + p_{T_{Z}}^{2}}$  is
applied to the HERWIG gluon radiation.
\end{thebibliography}
\end{document}